# How does boiling occur in lattice Boltzmann simulations?

Q. Li[*], Y. Yu, and Z. X. Wen

*School of Energy Science and Engineering, Central South University, Changsha 410083, China*

*Corresponding author: qingli@csu.edu.cn

**Abstract**

In recent years, the lattice Boltzmann (LB) method has been widely employed to simulate boiling phenomena [A. Márkus and G. Házi, Phys. Rev. E 83, 046705 (2011); Biferale *et al.*, Phys. Rev. Lett. 108, 104502 (2012); Li *et al.*, Phys. Rev. E 96, 063303 (2017); Wu *et al.*, Int. J. Heat Mass Transfer 126, 773 (2018)]. However, a very important issue still remains open, i.e., how does boiling occur in the LB simulations? For instance, the existing LB studies showed that the boiling on a hydrophobic surface begins at a lower wall superheat than that on a hydrophilic surface, which qualitatively agrees well with experimental studies, but no one has yet explained how this phenomenon appears in the LB simulations and what happened in the simulations after changing the wettability of the heating surface. In this paper, the LB boiling mechanism is revealed by analyzing boiling on a flat surface with mixed wettability and boiling on a structured surface with homogeneous wettability. Through a theoretical analysis, we demonstrate that, when the same wall superheat is applied, in the LB boiling simulations the fluid density near the heating surface decreases faster on a hydrophobic surface than that on a hydrophilic surface. Accordingly, a lower wall superheat can induce the phase transition from liquid to vapor on a hydrophobic surface than that on a hydrophilic surface. Furthermore, a similar theoretical analysis shows that the fluid density decreases fastest at concave corners in the case of a structured surface with homogeneous wettability, which explains why vapor bubbles are nucleated at concave corners in the LB simulations of boiling on structured surfaces.

# I. Introduction

Boiling is commonly encountered in daily life and plays a very important role in a wide range of technological and industrial applications such as nuclear reactors, heat exchangers, refrigeration, and cooling of high-power electronics [1-3]. As a very efficient mode of heat transfer, boiling is an extremely complex process associated with various physical components, such as the nucleation, growth, and departure of vapor bubbles, the transport of latent heat with the liquid-vapor phase transition, the instability of liquid-vapor interfaces, and the hydrodynamic interactions between bubbles [4]. In the past decades, boiling heat transfer has been extensively studied by experiments and many different types of empirical correlations predicting the heat transfer coefficient of boiling have been proposed with some adjustable parameters [5]. However, owing to the extreme complexity of boiling phenomena, most studies are empirical in nature and the fundamentals of boiling have not been well understood.

In recent years, with the rapid development of high-performance computational technology, it has been the belief of many researchers that numerical simulations of the complete boiling process may be an option for understanding the boiling fundamentals [1, 4]. Up to date, various attempts have been made to investigate boiling heat transfer numerically and the existing numerical methods for boiling simulations can be generally classified into the following categories.

(i) *Molecular method*. The molecular dynamics (MD) method is a computational method of studying the physical movements of atoms and molecules in the context of N-body simulation. In the MD simulations, the trajectories of particles are determined by numerically solving Newton's equations of motion [5]. By choosing reasonable interactions between the particles and using proper boundary conditions, the boiling of a liquid can be reproduced by the MD method in a natural way without introducing numerical artifacts [6]. Therefore, it is a very promising method to grasp the fundamental physics of boiling phenomena. However, the existing MD simulations of boiling heat transfer are usually limited to systems of very small size.

(ii) *Continuum method with empirical models*. This category includes the Level-Set method [7, 8],

the Volume-of-Fluid method [9], the Front-Tracking method [10], etc. These interface-capturing/tracking methods are very popular and successful in simulating isothermal two-phase flows and have been extended to simulate boiling phenomena by employing an empirical model for phase change as well as *ad hoc* assumptions on the contact line dynamics. A brief review of these methods for simulating boiling heat transfer can be found in Ref. [11]. The major drawback of these methods is that in the simulations of nucleate boiling the nucleation sites are prescribed *a priori* on the heating surface [12] and therefore they cannot provide any insights regarding the bubble nucleation.

(iii) *Continuum method with thermodynamics*. In recent years, a diffuse-interface method based on the dynamic van der Waals theory [13, 14] has also been applied to simulate boiling heat transfer [15], which directly solves the van der Waals-Navier-Stokes equations and provides an alternative continuum approach for investigating boiling phenomena. Owing to the use of diffuse interfaces, the stress and thermal singularities at the contact line are resolved automatically [16]. Moreover, the dynamic van der Waals theory treats the evaporation or condensation rate at the liquid-vapor interface as an outcome of calculation [4, 17] rather than a prerequisite as in the continuum method with empirical models.

(iv) *Lattice Boltzmann method*. Besides the aforementioned categories, the lattice Boltzmann (LB) method [18-21], which is a mesoscopic method built on the kinetic Boltzmann equation, has been widely employed to simulate boiling heat transfer in the past decade. Particularly, most of these studies are based on the pseudopotential multiphase LB method [22, 23], in which the fluid interactions are modeled by an interparticle potential and the phase separation is achieved by imposing a short-range attraction between different phases. Hence an important advantage of the pseudopotential LB method for simulating multiphase flows lies in that the interface between different phases can arise, deform, and migrate naturally without using any techniques to track or capture the interface [21], which is required in the aforementioned interface-capturing/tracking methods. Moreover, the implementation of wetting boundaries (contact angles) in the pseudopotential multiphase LB method [24, 25] is much simpler than that in conventional numerical methods.

Historically, in the LB community the first study of boiling phenomena was conducted by Zhang and Chen [26]. They successfully reproduced a nucleate boiling process by employing a standard Rayleigh-Bénard setup, in which the upper and lower solid walls obey the no-slip boundary condition and the horizontal boundary condition is periodic. Subsequently, Márkus and Házi [27, 28] investigated a heterogeneous boiling process on a heated plate with a square cavity. By applying various geometrical configurations for the heated plate, the bubble departure diameter and release period were studied. Later, Biferale *et al*. [29] carried out a three-dimensional LB study of nucleate boiling and demonstrated that the pseudopotential multiphase LB method with a non-ideal thermodynamic pressure tensor can lead to a consistent definition of latent heat with the Clausius-Clapeyron relation being satisfied. In 2015, Li *et al*. [30] and Gong and Cheng [31] successfully simulated three boiling stages of pool boiling, i.e., nucleate boiling, transition boiling, and film boiling, and reproduced a boiling curve based on the pseudopotential multiphase LB method.

Furthermore, inspired by some recent experimental studies of boiling on surfaces with mixed wettability [32-34], Cheng *et al*. [35, 36] and Li *et al*. [37, 38] have conducted LB simulations of boiling heat transfer on surfaces with mixed wettability. For instance, Gong and Cheng [35] studied the boiling heat transfer on flat surfaces with mixed wettability and found that adding hydrophobic spots on smooth hydrophilic surfaces can promote bubble nucleation and reduce the nucleation time. Li *et al*. [37] investigated the boiling heat transfer performance on a type of pillar-textured surface with mixed wettability. The pillars are composed of hydrophilic side walls and hydrophobic tops. They found that increasing the contact angle of the tops of pillars leads to a leftward shift of the boiling curve and a leftward-upward shift of the heat transfer coefficient curve. Besides, Ma and Cheng [39] recently studied the boiling heat transfer on micro-pillar and micro-cavity hydrophilic heaters, and Wu *et al*. [40] investigated co-existing boiling and condensation in a confined micro-space using the pseudopotential multiphase LB method.

Although considerable efforts have been made in applying the LB method to simulate boiling heat transfer, little attention has been devoted to investigating the *LB boiling mechanism*, i.e., the boiling mechanism in the LB simulations. Therefore a significant issue still remains open, i.e., how does boiling occur in the LB simulations? For example, the existing LB studies have shown that the onset of boiling on a hydrophobic surface requires a lower wall superheat than that on a hydrophilic surface [35, 37], which is qualitatively in good agreement with experimental studies [41-43], but no one has explained how this phenomenon appears in the LB simulations and what happened in the simulations after changing the wettability of the heating surface. The aim of the present work is therefore to investigate the LB boiling mechanism and figure out the rationale behind the boiling phenomena produced in the LB simulations. Similar to most of the existing LB studies on boiling heat transfer, the present work is also based on the pseudopotential multiphase LB method. Specifically, the LB boiling mechanism will be investigated by simulating and analyzing boiling on a flat surface with mixed wettability and boiling on a structured surface with homogeneous wettability. The rest of the present paper is organized as follows. The pseudopotential multiphase LB method is briefly introduced in Sec. II. Subsequently, numerical investigation and theoretical analyses of the LB boiling mechanism are presented in Sec. III. Finally, a brief summary is provided in Sec. IV.

## II. The pseudopotential LB method for non-ideal fluids

The original pseudopotential multiphase LB method was proposed by Shan and Chen [22, 23] around 1993. They introduced an interparticle potential $\psi(\mathbf{x})$ (i.e., the pseudopotential) to mimic the interactions between the particles on the nearest-neighboring sites. For single-component non-ideal fluids, the interaction force acting on the particles at site $\mathbf{x}$ is given by [44]

$$\mathbf{F}_\mathrm{m}(\mathbf{x},t) = -G\psi(\mathbf{x}) \sum_\alpha w_\alpha \psi(\mathbf{x}+\mathbf{e}_\alpha \delta_t) \mathbf{e}_\alpha, \tag{1}$$

where $G$ is a parameter that controls the strength of the interaction force, $w_\alpha$ are the weights of the interaction force, $\mathbf{e}_\alpha$ is the discrete velocity in the $\alpha$ th direction, $t$ is the time, and $\delta_t$ is the time step. For the two-dimensional nine-velocity (D2Q9) lattice model, the weights $w_\alpha$ are given by $w_\alpha = 1/3$ for $|\mathbf{e}_\alpha|^2 = 1$ and $w_\alpha = 1/12$ for $|\mathbf{e}_\alpha|^2 = 2$, respectively [45]. As highlighted by Succi [46], the magic of the pseudopotential interaction force is that it not only gives a non-monotonic equation of state supporting phase transition but also yields non-zero surface tension. Hence in the pseudopotential LB method the phase segregation between different phases can emerge automatically as a result of the particle interactions. Over the past two and a half decades, significant progress has been achieved in the pseudopotential multiphase LB method and a comprehensive review regarding the theory and applications of this method can be found in Ref. [21].

In the literature, the LB equation with a multiple-relaxation-time (MRT) collision operator [47, 48] has been demonstrated to be superior over the LB equation with the Bhatnagar-Gross-Krook (BGK) collision operator [49, 50] in terms of numerical stability, but in the present work the LB-BGK equation is adopted based on the consideration that theoretical analyses can be conducted much more easily when using the LB-BGK equation and the conclusions of our work are not affected by employing either the MRT or the BGK collision operator. Generally, the LB-BGK equation can be written as follows:

$$f_\alpha(\mathbf{x} + \mathbf{e}_\alpha \delta_t, t + \delta_t) = f_\alpha(\mathbf{x}, t) - \frac{1}{\tau}\left[f_\alpha(\mathbf{x}, t) - f_\alpha^{eq}(\mathbf{x}, t)\right] + \delta_t F_\alpha(\mathbf{x}, t), \tag{2}$$

where $f_\alpha$ is the density distribution function, $\tau$ is the non-dimensional relaxation time, $F_\alpha$ is the forcing term, and $f_\alpha^{eq}$ is the equilibrium distribution function given by

$$f_\alpha^{eq} = \omega_\alpha \rho \left[1 + \frac{\mathbf{e}_\alpha \cdot \mathbf{v}}{c_s^2} + \frac{(\mathbf{e}_\alpha \cdot \mathbf{v})^2}{2c_s^4} - \frac{(\mathbf{v} \cdot \mathbf{v})}{2c_s^2}\right], \tag{3}$$

where $c_s = c/\sqrt{3}$ is lattice sound speed, in which $c = 1$ is the lattice speed, and $\omega_\alpha$ are the weights for the equilibrium distribution function. For the D2Q9 lattice model, the weights $\omega_\alpha$ are given by $\omega_0 = 4/9$ for $\mathbf{e}_\alpha = 0$, $\omega_\alpha = 1/9$ for $|\mathbf{e}_\alpha|^2 = 1$, and $\omega_\alpha = 1/36$ for $|\mathbf{e}_\alpha|^2 = 2$, respectively [49].

The force of the system is incorporated into the LB equation through the forcing term in Eq. (2). In the LB community, the forcing scheme of Guo *et al*. [51] is widely applied, which is given by

$$F_\alpha(\mathbf{x}, t) = \omega_\alpha \left(1 - \frac{1}{2\tau}\right)\left[\frac{(\mathbf{e}_\alpha - \mathbf{v})}{c_s^2} + \frac{(\mathbf{e}_\alpha \cdot \mathbf{v})}{c_s^4}\mathbf{e}_\alpha\right] \cdot \mathbf{F}. \tag{4}$$

Correspondingly, the macroscopic density $\rho$ and velocity $\mathbf{v}$ are defined as

$$\rho = \sum_\alpha f_\alpha, \quad \rho\mathbf{v} = \sum_\alpha \mathbf{e}_\alpha f_\alpha + \frac{\delta_t}{2}\mathbf{F}, \tag{5}$$

where $\mathbf{F}$ is the total force exerted on the system, including the pseudopotential interaction force $\mathbf{F}_m$ and the buoyant force $\mathbf{F}_b$.

To incorporate a non-ideal equation of state into the pseudopotential multiphase LB method, the pseudopotential $\psi(\mathbf{x})$ can be chosen as $\psi(\mathbf{x}) = \sqrt{2(p_{EOS} - \rho c_s^2)/Gc^2}$ [52-54]. Here $p_{EOS}$ is the non-ideal equation of state. When the pseudopotential takes such a square-root form, the coexisting liquid and vapor densities ($\rho_L$ and $\rho_V$) produced by the pseudopotential LB method are usually inconsistent with those given by the Maxwell construction [55]. Such inconsistency can be approximately eliminated by adjusting the mechanical stability condition of the pseudopotential LB method through an improved forcing scheme [48, 54], but note that this issue does not affect the LB boiling mechanism investigated in the present work, and therefore we still employ the forcing scheme of Guo *et al.* [51], which can simplify the theoretical analyses in the subsequent section.

In order to take the liquid-vapor phase transition into account, the above formulations should be combined with a temperature solver. In 2002, He and Doolen [53] investigated the thermodynamic foundations of kinetic theory for non-ideal fluids and derived an energy transport equation from a kinetic equation that combines Enskog's theory for dense fluids and the mean-field theory for long-range molecular interactions. Similarly, a general equation for the total energy density of non-ideal fluids can be found in the work of Onuki [13], which can be transformed to the following equation for the internal energy density of non-ideal fluids:

$$\partial_t \hat{e} + \nabla \cdot (\mathbf{v}\hat{e}) = \nabla \cdot (\lambda \nabla T) + (\mathbf{\Phi} - \mathbf{\Pi}) : \nabla \mathbf{v}, \tag{6}$$

where $\hat{e} = \rho e_n$ is the internal energy density [14] ($e_n$ is the internal energy of non-ideal fluids), $\lambda$ is the thermal conductivity, $\mathbf{\Phi}$ is the dissipative stress tensor, and $\mathbf{\Pi} = p_{EOS}\mathbf{I} + \mathbf{T}$ is the nonviscous stress, in which $\mathbf{I}$ is the unit tensor and $\mathbf{T}$ is the contribution to the pressure tensor depending on density gradients [56]. According to thermodynamics, the following thermodynamic relation can be obtained:

$$\mathrm{d}e_\mathrm{n} = c_V \mathrm{d}T + \left[ T \left( \frac{\partial p_\mathrm{EOS}}{\partial T} \right)_V - p_\mathrm{EOS} \right] \mathrm{d}V , \qquad (7)$$

where $V = 1/\rho$ and $c_V$ is the specific heat at constant volume. Using the above relation as well as the continuity equation $\mathrm{D}\rho/\mathrm{D}t = -\rho \nabla \cdot \mathbf{v}$, Eq. (6) can be transformed to

$$\rho c_V \left( \partial_t T + \mathbf{v} \cdot \nabla T \right) = \nabla \cdot (\lambda \nabla T) - T \left( \frac{\partial p_\mathrm{EOS}}{\partial T} \right)_\rho \nabla \cdot \mathbf{v} + (\mathbf{\Phi} - \mathbf{T}) : \nabla \mathbf{v} . \qquad (8)$$

In most of the existing LB studies about boiling phenomena, the pseudopotential multiphase LB method is combined with a temperature solver for Eq. (8), either a thermal LB equation [27-29, 35, 57] or a finite-difference algorithm [30, 37, 38, 40]. The last term on the right-hand side of Eq. (8) is usually neglected in these studies. In the present work, a finite-difference algorithm is adopted and its details can be found in Ref. [30].

### III. Numerical investigation and theoretical analyses

In this section, numerical investigation and theoretical analyses are conducted to study the LB boiling mechanism. Two types of heating surfaces are considered, i.e., a flat surface with mixed wettability and a structured surface with homogeneous wettability. The Peng-Robinson equation of state is utilized in the present study, i.e., [52]

$$p_\mathrm{EOS} = \frac{\rho RT}{1 - b\rho} - \frac{a \varphi(T) \rho^2}{1 + 2b\rho - b^2 \rho^2} , \qquad (9)$$

where $\varphi(T) = \left[ 1 + \left( 0.37464 + 1.54226\omega - 0.26992\omega^2 \right) \left( 1 - \sqrt{T/T_c} \right) \right]^2$, $R$ is the gas constant, $a = 0.45724 R^2 T_c^2 / p_c$, and $b = 0.0778 R T_c / p_c$. The parameter $\omega = 0.344$ is the acentric factor and the subscript "c" represents the critical parameters. In our simulations, $a = 3/49$, $b = 2/21$, and $R = 1$ [30], which yields $T_c \approx 0.10938$. The saturation temperature is chosen as $T_0 = 0.86 T_c$, which corresponds to $\rho_L \approx 6.45$ and $\rho_V \approx 0.144$ when using the forcing scheme of Guo *et al.* [51]. The thermal conductivity $\lambda$ is chosen to be proportional to the density, i.e., $\lambda = \rho \chi c_V$, in which $\chi = 0.021$ is the thermal diffusivity. Similar to some previous studies [26, 27, 30], the present study also employs a constant $c_V$, which is taken as $c_V = 6$. The quantities in the present paper are all based on the lattice units, i.e., the units of the LB method.

### A. Boiling on a flat surface with mixed wettability

#### *1. Numerical simulation*

The test of boiling on a flat surface with mixed wettability is firstly considered, which is illustrated in Fig. 1, where "A" and "B" represent the hydrophobic and hydrophilic stripes, respectively. The width of the hydrophobic stripe is equal to that of the hydrophilic stripe. In our simulations, the grid system is chosen as $L_x \times L_y = 295 \text{ l.u.} \times 300 \text{ l.u.}$ Here l.u. denotes lattice units. The flat surface is placed at the bottom of the computational domain and the width of the stripes is taken as 49 l.u. The halfway bounce-back scheme [58] is applied for the no-slip boundary condition, while the periodic boundary condition is employed in the horizontal direction.

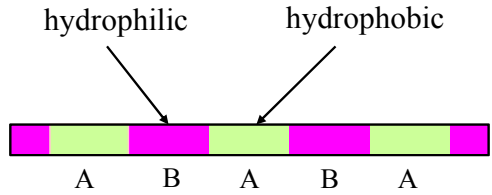

**Fig. 1**. Sketch of a flat surface with mixed wettability. "A" and "B" represent the hydrophobic and hydrophilic stripes, respectively. The periodic boundary condition is employed in the horizontal direction.

Initially, the liquid ($y < 2L_y/3$) is placed below its vapor with a diffuse interface. The temperature of the heating surface is taken as $T_w = T_0 + \Delta T$, where $\Delta T$ is the wall superheat. The intrinsically equilibrium contact angles of the hydrophilic and hydrophobic stripes are chosen as $\theta \approx 79°$ and $\theta \approx 121°$, respectively, and the gravitational acceleration is given by $g = 4 \times 10^{-5}$. The contact angles are implemented by specifying the pseudopotentials of solid nodes via virtual densities and the details can be found in Ref. [25]. Three cases are modeled in our simulations, i.e., $\Delta T = 0.04T_c$, $0.095T_c$, and $0.135T_c$, respectively. Some snapshots of the density contours of these cases are displayed in Fig. 2 with the kinematic viscosity being taken as $v = 0.25$, which corresponds to $\tau = 1.25$. As shown in the figure, there is no boiling at a low wall superheat. When the wall superheat increases, nucleate boiling can be observed in the case of $\Delta T = 0.095T_c$, while the case of $\Delta T = 0.135T_c$ leads to film boiling. Figure 3 shows the snapshots of boiling process at $\Delta T = 0.095T_c$ with $v = 0.15$ (i.e., $\tau = 0.95$). By comparing Fig. 2(b) with Fig. 3, we can find that the bubble growth and departure are speeded up when decreasing the relaxation time (i.e., increasing the Rayleigh number).

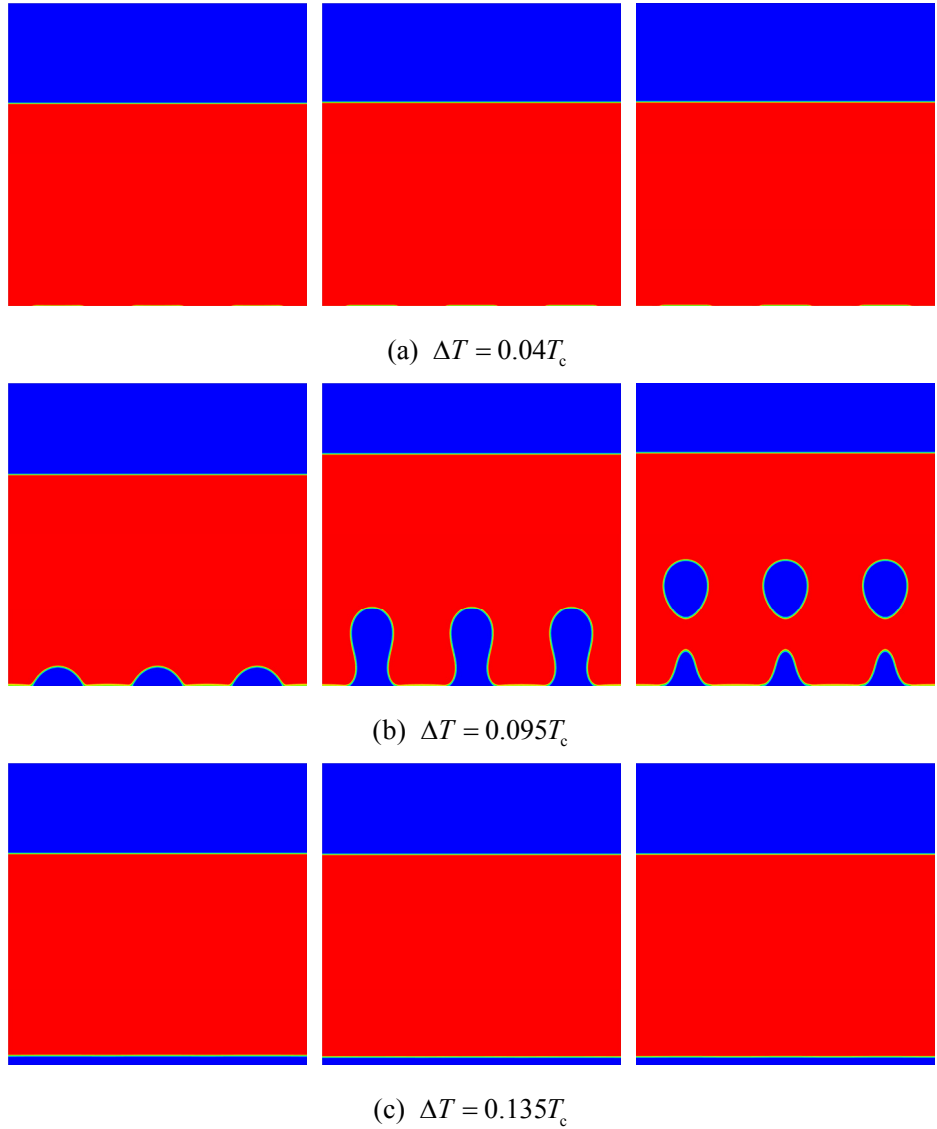

(a) $\Delta T = 0.04T_c$

(b) $\Delta T = 0.095T_c$

(c) $\Delta T = 0.135T_c$

**Fig. 2**. Simulation of boiling on a flat surface with mixed wettability at $\Delta T = 0.04T_c$, $0.095T_c$ and $0.135T_c$ with $v = 0.25$. From left to right: $t = 12000\delta_t$, $36000\delta_t$, and $48000\delta_t$, respectively.

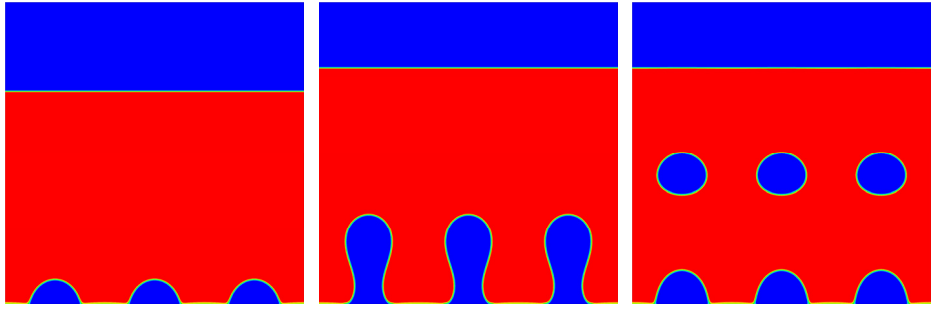

**Fig. 3**. Simulation of boiling on a flat surface with mixed wettability at $\Delta T = 0.095 T_c$ with $\nu = 0.15$. From left to right: $t = 12000 \delta_t$, $36000 \delta_t$, and $48000 \delta_t$, respectively.

Particularly, from the nucleate boiling processes shown in Fig. 2(b) and Fig. 3, it can be clearly seen that in the case of $\Delta T = 0.095 T_c$ vapor bubbles are nucleated at the hydrophobic stripes, whereas there is no bubble nucleation at the hydrophilic stripes. However, when the wall superheat is increased to $0.135 T_c$, the film boiling appears. In other words, the boiling on a hydrophobic surface begins at a lower wall superheat than that on a hydrophilic surface. Such a phenomenon is clear and consistent with those observed in experimental studies [41-43], molecular dynamics simulations [5], and previous LB studies [35, 37]. Nevertheless, as previously mentioned, no one has yet explained how this phenomenon occurs in the LB simulations and what happens when the wettability of the heating surface is changed in the LB boiling simulations.

## *2. Theoretical analysis and discussion*

To figure out the rationale behind the aforementioned boiling phenomenon produced in the LB simulations, a theoretical analysis is conducted to analyze the variation of the fluid density near the heating surface. Before doing this, we would like to make some statements about the theoretical analysis.

(i) Owing to the extreme complexity of boiling phenomena, it is very difficult or even impossible to perform a theoretical analysis for a general case. Therefore our analysis is focused on the variation of the density of the first grid layer near the heating surface at a very *early stage* of simulation, which means that the fluid velocity is neglected in the analysis, i.e., $\mathbf{v} = 0$.

(ii) Moreover, the periodic boundary condition is applied in the horizontal direction and the

non-dimensional relaxation time in the analysis is set to $\tau = 1$, through which the theoretical analysis can be considerably simplified.

(iii) The pseudopotential $\psi$ is plotted in Fig. 4 as a function of $\rho$ and $T$ by substituting the equation of state $p_{EOS} = p_{EOS}(\rho, T)$ into its expression. From the figure we can see that the pseudopotential increases as the density increases, but reduces with increasing the temperature.

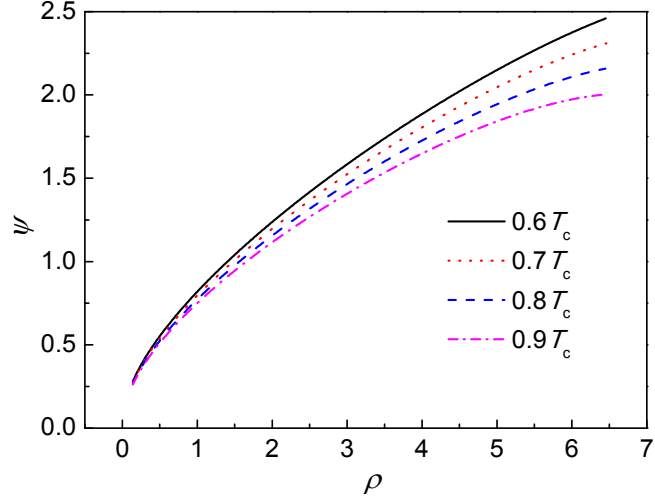

**Fig. 4**. Variations of the pseudopotential as a function of density and temperature.

By applying the aforementioned treatments, the collision process of the LB-BGK equation with the forcing term of Guo *et al.* [51] can be transformed to

$$f_\alpha^*(\mathbf{x}, t) = f_\alpha(\mathbf{x}, t) - \frac{1}{\tau}\left[f_\alpha(\mathbf{x}, t) - f_\alpha^{eq}(\mathbf{x}, t)\right] + \delta_t F_\alpha(\mathbf{x}, t)$$
$$= f_\alpha^{eq}(\mathbf{x}, t) + \frac{3}{2}\omega_\alpha(\mathbf{e}_\alpha \cdot \mathbf{F}), \quad (10)$$

where $\mathbf{e}_\alpha \cdot \mathbf{F} = e_{\alpha x}F_x + e_{\alpha y}F_y$. Note that $c_s^2 = 1/3$ and $\delta_t = 1$ are used in the above transformation. The streaming process is given by $f_\alpha(\mathbf{x} + \mathbf{e}_\alpha \delta_t, t + \delta_t) = f_\alpha^*(\mathbf{x}, t)$.

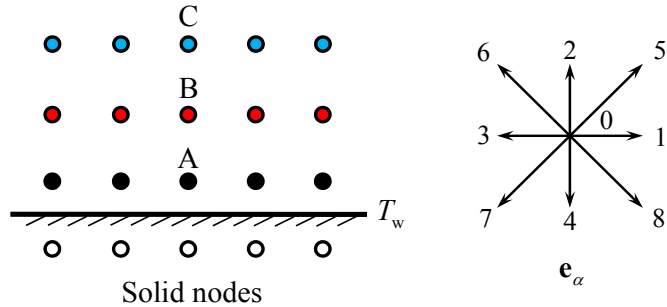

**Fig. 5**. Sketch of grid layers near a flat surface.

Figure 5 depicts three grid layers near a flat surface and a ghost layer of solid nodes underneath the flat surface, which is located at the middle of the fluid boundary layer (i.e., the layer of A) and the ghost layer because of using the halfway bounce-back scheme [58]

$$f_{\bar{\alpha}}(A, t+\delta_t) = f_{\alpha}^{*}(A, t), \tag{11}$$

where $f_{\bar{\alpha}}$ is the unknown distribution function in the $\bar{\alpha}$ th direction ($\mathbf{e}_2$, $\mathbf{e}_5$, and $\mathbf{e}_6$ in Fig. 5) and $\mathbf{e}_{\bar{\alpha}} = -\mathbf{e}_{\alpha}$. According to Eqs. (10) and (11), we can obtain

$$f_2(A, t+\delta_t) = f_4^{*}(A, t) = f_4^{eq}(A, t) + \frac{3}{2}\omega_4\left[-F_y(A, t)\right], \tag{12}$$

$$f_5(A, t+\delta_t) = f_7^{*}(A, t) = f_7^{eq}(A, t) + \frac{3}{2}\omega_7\left[-F_x(A, t) - F_y(A, t)\right], \tag{13}$$

$$f_6(A, t+\delta_t) = f_8^{*}(A, t) = f_8^{eq}(A, t) + \frac{3}{2}\omega_8\left[F_x(A, t) - F_y(A, t)\right]. \tag{14}$$

The distribution functions in other directions can be obtained according to the streaming process, i.e.,

$$f_0(A, t+\delta_t) = f_0^{*}(A, t) = f_0^{eq}(A, t), \tag{15}$$

$$f_1(A, t+\delta_t) = f_1^{*}(A, t) = f_1^{eq}(A, t) + \frac{3}{2}\omega_1\left[F_x(A, t)\right], \tag{16}$$

$$f_3(A, t+\delta_t) = f_3^{*}(A, t) = f_3^{eq}(A, t) + \frac{3}{2}\omega_3\left[-F_x(A, t)\right], \tag{17}$$

$$f_4(A, t+\delta_t) = f_4^{*}(B, t) = f_4^{eq}(B, t) + \frac{3}{2}\omega_4\left[-F_y(B, t)\right], \tag{18}$$

$$f_7(A, t+\delta_t) = f_7^{*}(B, t) = f_7^{eq}(B, t) + \frac{3}{2}\omega_7\left[-F_x(B, t) - F_y(B, t)\right], \tag{19}$$

$$f_8(A, t+\delta_t) = f_8^{*}(B, t) = f_8^{eq}(B, t) + \frac{3}{2}\omega_8\left[F_x(B, t) - F_y(B, t)\right]. \tag{20}$$

Here it should be noted that $f_1(A, t+\delta_t)$ is theoretically given by $f_1^{*}(A - \mathbf{e}_1\delta_t, t)$, but the fluid nodes in the layer of A have the same properties owing to using the periodic boundary condition in the horizontal direction, which means $f_1^{*}(A - \mathbf{e}_1\delta_t, t) = f_1^{*}(A, t)$. The directions of $\mathbf{e}_3$, $\mathbf{e}_4$, $\mathbf{e}_7$, and $\mathbf{e}_8$ are treated in a similar way. By combining Eqs. (12)-(14) with Eqs. (15)-(20), we can obtain

$$\rho(A, t+\delta_t) = \sum_{\alpha} f_{\alpha}(A, t+\delta_t)$$
$$= f_0^{eq}(A, t) + f_1^{eq}(A, t) + f_3^{eq}(A, t) + f_4^{eq}(A, t) + f_7^{eq}(A, t) + f_8^{eq}(A, t)$$
$$+ f_4^{eq}(B, t) + f_7^{eq}(B, t) + f_8^{eq}(B, t) + \frac{3}{2}(\omega_4 + \omega_7 + \omega_8)\left[-F_y(A, t) - F_y(B, t)\right]. \tag{21}$$

Substituting Eq. (3) as well as the weights $\omega_{\alpha}$ into Eq. (21) yields

$$\rho(A, t+\delta_t) = \frac{5}{6}\rho(A,t) + \frac{1}{6}\rho(B,t) - \frac{1}{4}\left[F_y(A,t) + F_y(B,t)\right]. \tag{22}$$

This equation serves as our basis for analyzing the variation of the density of the first grid layer near the flat heating surface.

A representative analysis can be made from $t=0$ to $\delta_t$ since we have $\rho(A) = \rho(B) = \rho_L$ and $F_y(B) = 0$ at the initial time $t=0$, which leads to

$$\rho(A, \delta_t) = \rho_L - \frac{1}{4}F_y(A, 0), \tag{23}$$

Neglecting the buoyant force, the following equation can be obtained according to Eq. (1) and Fig. 5:

$$F_y(A, 0) = -\frac{G}{2}\psi(A)\left[\psi(B) - \psi(S)\right], \tag{24}$$

where $\psi(S)$ is the pseudopotential of solid nodes. When the pseudopotential takes the square-root form, namely $\psi(\mathbf{x}) = \sqrt{2(p_{EOS} - \rho c_s^2)/Gc^2}$, the parameter $G$ is usually chosen as $G = -1$ so as to ensure that the whole term in the square root is positive [52]. Combining Eq. (23) with Eq. (24) and $G = -1$, we can obtain

$$\rho(A, \delta_t) = \rho_L - \frac{1}{8}\psi(A)\left[\psi(B) - \psi(S)\right]. \tag{25}$$

According to Fig. 4, we can conclude that $\psi(B) > \psi(S)$ because $\rho(B) = \rho_L$ and $T(B) = T_0$ at the initial time, while the virtual density of solid nodes $\rho(S)$ is smaller than $\rho_L$ [25] and its virtual temperature is $T(S) = T_w > T_0$. Hence $\rho(A, \delta_t)$ will be decreased in comparison with $\rho_L$.

Now it is clear that the surface wettability in the LB simulations definitely affects the boiling phenomena since the pseudopotential of solid nodes, namely $\psi(S)$ in Eq. (25), varies with the contact angle [25]. Moreover, for a hydrophobic surface and a hydrophilic surface with the same wall superheat, the following equation can be obtained from Eq. (25):

$$\rho_{pho}(A, \delta_t) - \rho_{phi}(A, \delta_t) = \frac{1}{8}\psi(A)\left[\psi(S)_{pho} - \psi(S)_{phi}\right], \tag{26}$$

where "pho" and "phi" represent the cases of hydrophobic and hydrophilic surfaces, respectively. When the same wall superheat is applied, from Fig. 4 it can be readily found that $\psi(S)_{pho} < \psi(S)_{phi}$ because a larger contact angle corresponds to a smaller virtual density of solid nodes [25]. As a result, Eq. (26) yields $\rho_{\text{pho}}(A, \delta_t) < \rho_{\text{phi}}(A, \delta_t)$, which indicates that the fluid density near the heating surface decreases faster on a hydrophobic surface than that on a hydrophilic surface.

As times goes by, the force $F_y(B, t)$ in Eq. (22) will be no longer zero. Through a similar analysis, it can be found that $F_y(B, t) > 0$ at the early stage of simulation, which means that the force $F_y(B, t)$ will also contribute to a decrease of the density $\rho(A, t + \delta_t)$, as shown in Eq. (22). Finally, we would like to explain why a considerable decrease of the fluid density near the heating surface may lead to the liquid-vapor phase transition. In fact, similar discussions can be found in some studies of non-ideal equations of state [59] and molecular dynamics simulations of boiling [60, 61]. A phase diagram with a coexistence curve and a spinodal line is sketched in Fig. 6, where the solid and dashed lines represent the coexistence curve and the spinodal line, respectively. The left and right parts of the coexistence curve denote the saturated vapor line and the saturated liquid line, respectively. These two lines meet at the critical point, forming a dome as shown in Fig. 6.

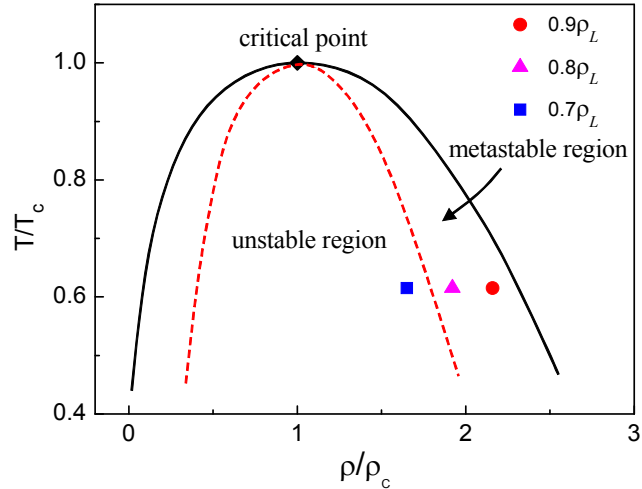

**Fig. 6**. Sketch of a phase diagram with coexistence curve (solid line) and spinodal line (dashed line).

The metastable region is located between the coexistence curve and the spinodal line. As illustrated in the literature [60], the liquid-vapor phase transition may not occur if the fluid density deviates slightly from the saturation liquid density $\rho_L$, such as the cases denoted by the solid circle and triangle in Fig. 6, which are located in the metastable region and correspond to $\rho = 0.9\rho_L$ and $0.8\rho_L$, respectively. In contrast, when the fluid density is further decreased and enters into the unstable region, such as the case denoted by the solid square in Fig. 6, the phase transition can be triggered owing to the unstable feature of the region [60]. This explains why a faster decrease of the fluid density near the heating surface results in a earlier liquid-vapor phase transition.

### B. Boiling on a structured surface with homogeneous wettability

In this section, we turn our attention to boiling on a structured heating surface with homogeneous wettability. The simulation parameters and conditions are basically the same as those in the preceding section except that the wettability of the structured surface is intrinsically homogeneous ($\theta \approx 79°$) rather than hydrophobic-hydrophilic mixed. The grid system is chosen as $L_x \times L_y = 151 \text{ l.u.} \times 300 \text{ l.u.}$. The heating surface is placed at the bottom of the domain and its height is 30 l.u. A square cavity is located at the center of the heating surface with its width and height of 45 l.u. and 25 l.u., respectively. Initially, the liquid ($y < 2L_y/3$) is placed below its vapor. The halfway bounce-back scheme is applied at the heating surface and the periodic boundary condition is employed in the horizontal direction.

In this test, two cases are considered: $\Delta T = 0.04T_c$ and $0.095T_c$. Figures 7 and 8 display some snapshots of the density contours of the two cases with the non-dimensional relaxation time being taken as $\tau = 1.2$ and $1.0$, respectively. From the figures it can be seen that there is no boiling when applying a relatively low wall superheat, while nucleate boiling can be observed when the wall superheat is increased to $\Delta T = 0.095T_c$. Meanwhile, by comparing Fig. 8(b) with Fig. 7(b), it can be seen that the bubble growth and departure are relatively speeded up when decreasing the relaxation time. Moreover, from the left-hand panels of Figs. 7(b) and (8b), we can clearly see that two vapor bubbles are nucleated at the concave corners of the structured surface. As time elapses, these two bubbles coalesce and form a larger bubble filling the square cavity. Subsequently, the bubble grows up and gradually reaches its critical diameter. Then a vapor bubble will be detached from the heating surface, as shown in the right-hand panels of Figs. 7(b) and (8b). Such a phenomenon is well consistent with the classical theory of boiling heat transfer that bubble nucleation is very likely to occur at micro-cavities.

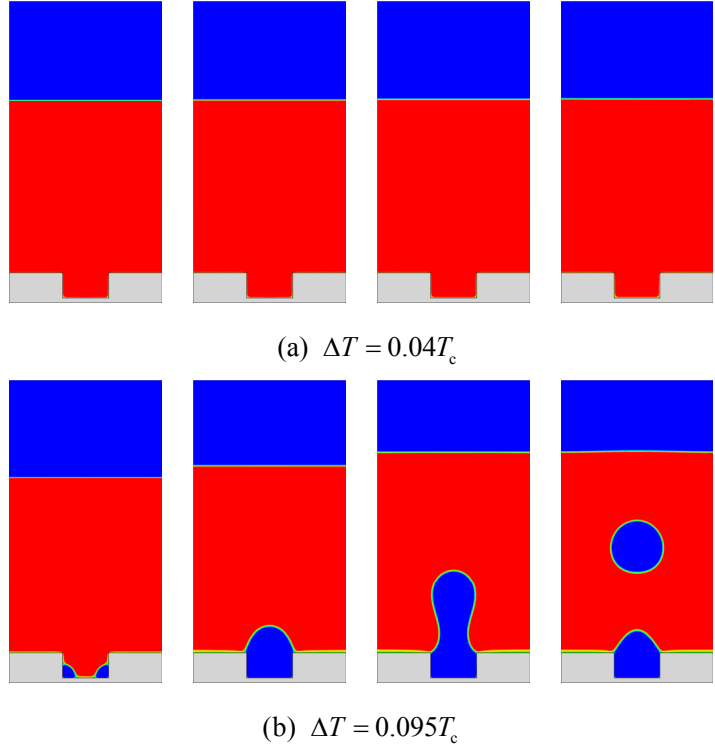

Fig. 7. Simulation of boiling on a structured surface with homogeneous wettability at $\Delta T = 0.04T_c$ and $0.095T_c$ with $\tau = 1.2$. From left to right: $t = 5\times10^3\delta_t$, $10^4\delta_t$, $2.5\times10^4\delta_t$, and $3.2\times10^4\delta_t$.

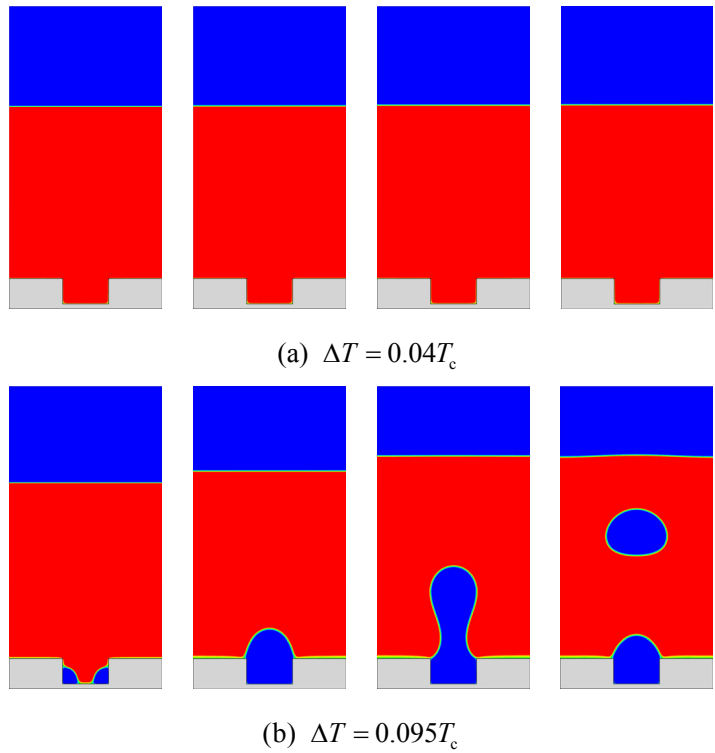

Fig. 8. Simulation of boiling on a structured surface with homogeneous wettability at $\Delta T = 0.04T_c$ and $0.095T_c$ with $\tau = 1$. From left to right: $t = 5\times10^3\delta_t$, $10^4\delta_t$, $2.5\times10^4\delta_t$, and $3.2\times10^4\delta_t$.

A theoretical analysis is now conducted to explain why the vapor bubbles are nucleated at the concave corners in the LB boiling simulations. By taking the left-hand concave corner as an example, Fig. 9 illustrates the grids around the corner. Due to using the halfway bounce-back scheme, the solid heating surface is located at the middle of fluid boundary nodes and solid nodes. The solid nodes have the same pseudopotential [25] so that the wettability of the heating surface is intrinsically homogeneous.

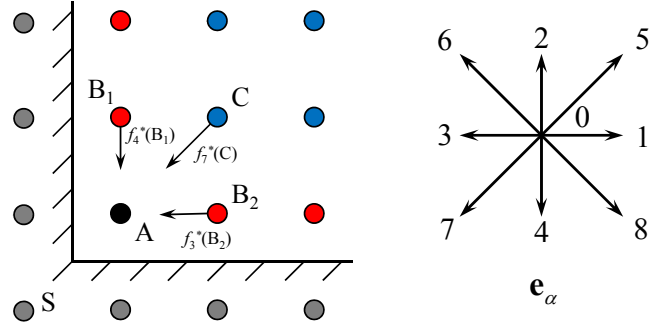

**Fig. 9**. Sketch of grids near a concave corner of the structured surface.

The statements made in the preceding section about the theoretical analysis are also applicable here. According to Eq. (10) and the halfway bounce-back scheme given by Eq. (11), we can obtain

$$f_1(A, t+\delta_t) = f_3^*(A, t) = f_3^{eq}(A, t) + \frac{3}{2}\omega_3\left[-F_x(A, t)\right], \tag{27}$$

$$f_2(A, t+\delta_t) = f_4^*(A, t) = f_4^{eq}(A, t) + \frac{3}{2}\omega_4\left[-F_y(A, t)\right], \tag{28}$$

$$f_5(A, t+\delta_t) = f_7^*(A, t) = f_7^{eq}(A, t) + \frac{3}{2}\omega_7\left[-F_x(A, t) - F_y(A, t)\right], \tag{29}$$

$$f_6(A, t+\delta_t) = f_8^*(A, t) = f_8^{eq}(A, t) + \frac{3}{2}\omega_8\left[F_x(A, t) - F_y(A, t)\right], \tag{30}$$

$$f_8(A, t+\delta_t) = f_6^*(A, t) = f_6^{eq}(A, t) + \frac{3}{2}\omega_6\left[-F_x(A, t) + F_y(A, t)\right]. \tag{31}$$

The distribution functions in other directions are obtained via the streaming process, i.e.,

$$f_0(A, t+\delta_t) = f_0^*(A, t) = f_0^{eq}(A, t), \tag{32}$$

$$f_3(A, t+\delta_t) = f_3^*(B_2, t) = f_3^{eq}(B_2, t) + \frac{3}{2}\omega_3\left[-F_x(B_2, t)\right], \tag{33}$$

$$f_4(A, t + \delta_t) = f_4^*(B_1, t) = f_4^{eq}(B_1, t) + \frac{3}{2}\omega_4\left[-F_y(B_1, t)\right], \tag{34}$$

$$f_7(A, t + \delta_t) = f_7^*(C, t) = f_7^{eq}(C, t) + \frac{3}{2}\omega_7\left[-F_x(C, t) - F_y(C, t)\right]. \tag{35}$$

Combining Eqs. (27)-(31) with Eqs. (32)-(35) yields

$$\begin{aligned}\rho(A, t + \delta_t) &= \sum_\alpha f_\alpha(A, t + \delta_t) \\ &= \frac{3}{4}\rho(A, t) + \frac{1}{9}\rho(B_1, t) + \frac{1}{9}\rho(B_2, t) + \frac{1}{36}\rho(C, t) - \frac{5}{24}F_x(A, t) \\ &\quad - \frac{5}{24}F_y(A, t) - \frac{1}{6}\left[F_y(B_1, t) + F_x(B_2, t)\right] - \frac{1}{24}\left[F_x(C, t) + F_y(C, t)\right].\end{aligned} \tag{36}$$

By comparing Eq. (36) with Eq. (22), it can be found that Eq. (36) is much more complicated as it involves the information of points A, $B_1$, $B_2$, and C in Fig. 9.

Nevertheless, a representative analysis can be performed in a similar way from $t = 0$ to $\delta_t$ since $\rho(A) = \rho(B_1) = \rho(B_2) = \rho(C) = \rho_L$ and $F_x(B_2) = F_y(B_1) = F_x(C) = F_y(C) = 0$ at the initial time. Correspondingly, the following equation can be obtained:

$$\rho(A, \delta_t) = \rho_L - \frac{5}{24}F_x(A, 0) - \frac{5}{24}F_y(A, 0). \tag{37}$$

According to Eq. (1) and $\psi(B_1) = \psi(B_2) = \psi(C)$ at $t = 0$, we can obtain

$$F_x(A, 0) = F_y(A, 0) = -\frac{5}{12}G\psi(A)\left[\psi(B) - \psi(S)\right], \tag{38}$$

where $B = B_1 = B_2$. Substituting Eq. (38) with $G = -1$ into Eq. (37) leads to

$$\rho(A, \delta_t) = \rho_L - \frac{25}{144}\psi(A)\left[\psi(B) - \psi(S)\right], \tag{39}$$

which shows that the density at the concave corner is decreased in comparison with $\rho_L$ because $\psi(B) > \psi(S)$, which has been analyzed in the preceding section.

For a fluid boundary node near the heating surface but far from the corners, the density can still be represented by Eq. (25). More importantly, according to Eq. and Eq. (25), we can find that

$$\rho_{\text{cor}}(A, \delta_t) - \rho_{\text{flat}}(A, \delta_t) = -\frac{7}{144}\psi(A)\left[\psi(B) - \psi(S)\right] < 0. \tag{40}$$

This equation indicates that, compared with the density of a fluid boundary node near the heating surface but far from the corners, the fluid density at a concave corner decreases much faster. By performing a similar analysis, it can be found that the fluid density at a convex corner is larger than that at a concave corner during the early stage of simulation. In other words, the fluid density decreases fastest at concave corners on a structured surface with homogeneous wettability, which explains why the vapor bubbles are nucleated at concave corners in the LB simulations of boiling on a structured surface.

As time elapses ($t > \delta_t$), the forces $F_y(B_1,t)$, $F_x(B_2,t)$, $F_x(C,t)$, and $F_y(C,t)$ in Eq. (36) are no longer zero. Nevertheless, it can be verified that these forces are all positive at the early stage of simulation (i.e., the nucleation process) according to their expressions and Fig. 4 on the basis of $\rho(A,t) < \rho(B,t) < \rho(C,t)$ and $T(A,t) > T(B,t) > T(C,t)$ during the nucleation process. To illustrate this point more clearly, the variations of these forces in the aforementioned LB boiling simulation ($\tau = 1$) are displayed in Fig. 10. From the figure we can observe that $F_y(B_1,t) \approx F_x(B_2,t)$ and $F_x(C,t) \approx F_y(C,t)$, which is attributed to a symmetry around the concave corner at the early stage of simulation (see Fig. 9). The buoyant force has been neglected since the gravitational acceleration is on the order of $10^{-5}$. With these positive forces, the density $\rho(A, t+\delta_t)$ is further decreased. Therefore it is no surprise that the bubbles are nucleated at the concave corners in the LB boiling simulations.

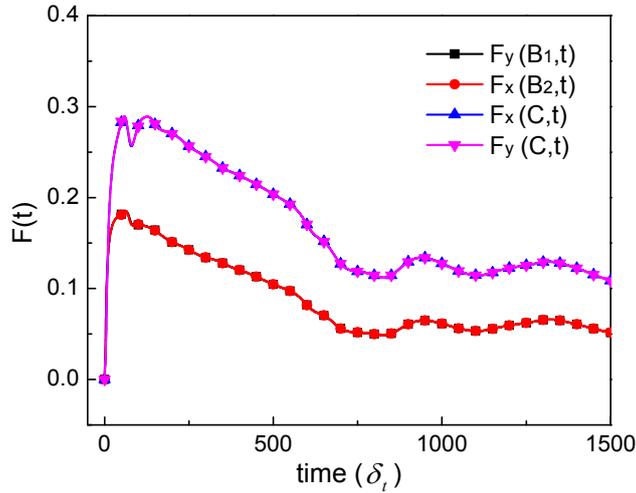

**Fig. 10**. Variations of the pseudopotential interaction forces at points $B_1$, $B_2$, and C during the early stage of the LB boiling simulation on a structured surface with homogeneous wettability.

## IV. Summary

Modeling boiling phenomena is very challenging in the community of computational fluid dynamics and various attempts have been made in the past decades. In recent years, the LB method, which is a mesoscopic numerical approach built on the kinetic Boltzmann equation, has also been applied to simulate boiling phenomena and has yielded some numerical results that are qualitatively in good agreement with experimental studies. However, in the previous LB studies very little attention has been paid to the LB boiling mechanism. In order to figure out the rationale behind the boiling phenomena produced in the LB simulations, in this paper we have investigated the LB boiling mechanism by analyzing boiling on a flat surface with mixed wettability and boiling on a structured surface with homogeneous wettability. Through a theoretical analysis, we have demonstrated that, when the same wall superheat is applied, the fluid density in the LB boiling simulations decreases faster on a hydrophobic surface than that on a hydrophilic surface. Therefore a lower wall superheat can induce the liquid-vapor phase transition on a hydrophobic surface than that on a hydrophilic surface. Moreover, by performing a similar theoretical analysis, we have shown that the fluid density decreases fastest at concave corners in the case of a structured surface with homogeneous wettability, which explains why vapor bubbles are nucleated at concave corners in the LB simulations of boiling on structured surfaces. To sum up, the present work has provided some insights into the LB boiling mechanism and we believe that the theoretical analyses as well as the related findings are pretty useful for promoting the fundamental understanding of boiling phenomena in the LB simulations.

## Acknowledgments

This work was supported by the National Natural Science Foundation of China (No. 51822606).

**Note:** In comparison with the 1st version, the present version has increased the thermal conductivity so as to ensure that the highest $T_w$ is smaller than $T_c$. Correspondingly, the related figures have been updated.